\documentclass[twoside,slac_one]{revtex4}
\usepackage{graphicx}
\usepackage{fancyhdr}
\usepackage{amsmath} 
\usepackage{bm}
\usepackage{amsxtra}
\usepackage{amssymb}
\usepackage{amsthm}
\usepackage{latexsym}
\usepackage{lscape}
\usepackage{rotating}
\begin{document}

\title{Early Neutrino Data in the NO$\nu$A Near Detector Prototype}

%

\author{M. Betancourt\footnote{For the NO$\nu$A Collaboration.}}
\affiliation{Department of Physics, University of Minnesota,
116 Church Street S.E., Minneapolis, Minnesota 55455, USA}

\begin{abstract}
NOvA is a long-baseline neutrino experiment using an off-axis
neutrino beam produced by the NuMI neutrino beam at Fermilab. The
NOvA experiment will study neutrino $\nu_{\mu}\rightarrow\nu_{e}$ oscillations. A short term goal for the NOvA experiment
is to develop a good understanding of the response of the detector.
These studies are being carried out with the full Near Detector
installed on the surface (NDOS) at Fermilab. This detector is currently
running and will acquire neutrino data for a year. Using beam muon
neutrino data, quasi-elastic charged-current interactions will be
studied. Status of the NDOS running and early data will be shown.

\end{abstract}

\maketitle

\thispagestyle{fancy}


\section{Introduction}
The NO$\nu$A experiment will study neutrino oscillations from $\nu_{\mu}$ to $\nu_e$ using the NuMI Off-Axis beam at Fermilab. The main goal of NO$\nu$A is to search for electron neutrino $\nu_e$ appearance. The NO$\nu$A detectors have been designed especially to identify electrons and provide good background rejection for selecting electron neutrino interactions. Observation of $\nu_{e}$ appearance allows for measurement of $\theta_{13}$, a search for the mass ordering, and a search for the CP violating phase $\delta$.  Secondary goals of the NO$\nu$A include improving the measurement of the $\nu_{\mu}$ disappearance parameters $\theta_{23}$ and $\Delta{m}^2_{32}$ and cross section measurements~\cite{zero}. \\
The NO$\nu$A experiment consist of two detectors, a near detector and a far detector, which will be located at different locations. The Near detector will be located at Fermilab and the far detector will be located 810 km away from the near at Ash River, Minnesota. The NO$\nu$A experiment will use a new type of detector, made of plastic PVC and liquid scintillator. The initial objective for the NO$\nu$A experiment is to obtain a good understanding of the detector for neutrino interactions. This is currently being achieved with a Near Detector prototype called (NDOS).  
\section{Near Detector On the Surface (NDOS)}
\begin{figure*}[h]
\centering
\includegraphics[width=40mm,height=40mm]{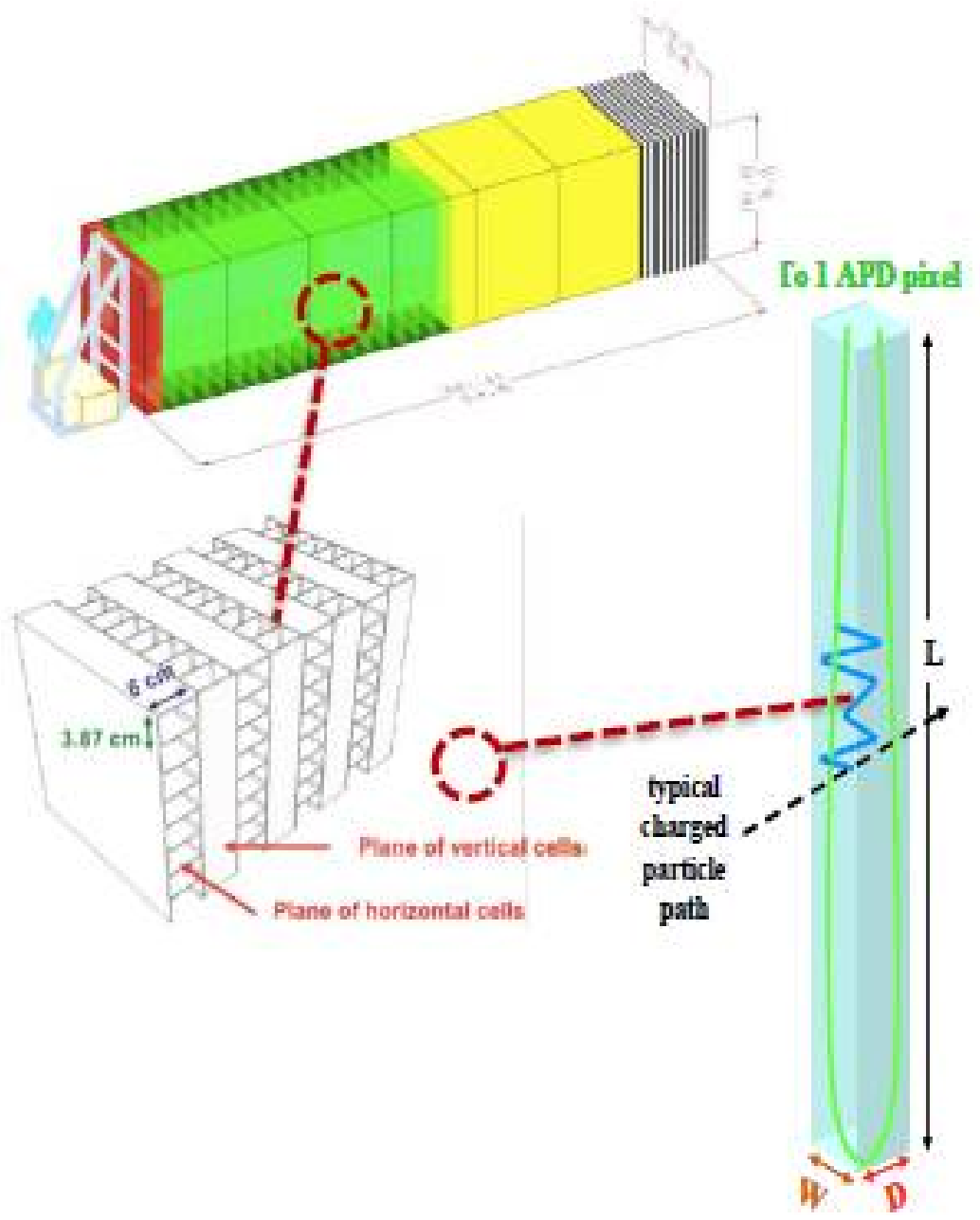}
\includegraphics[width=50mm,height=40mm]{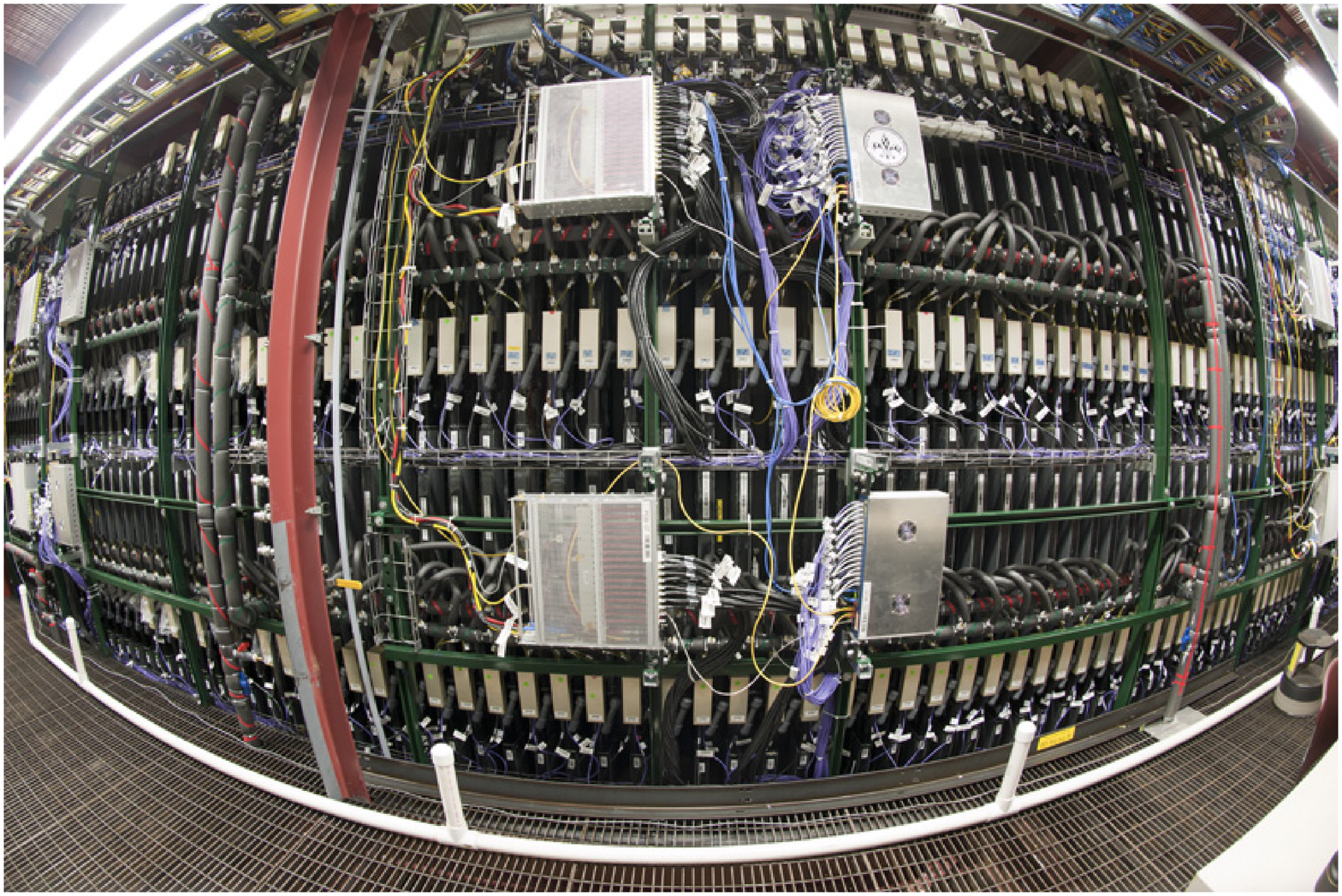}
\includegraphics[width=40mm,height=40mm]{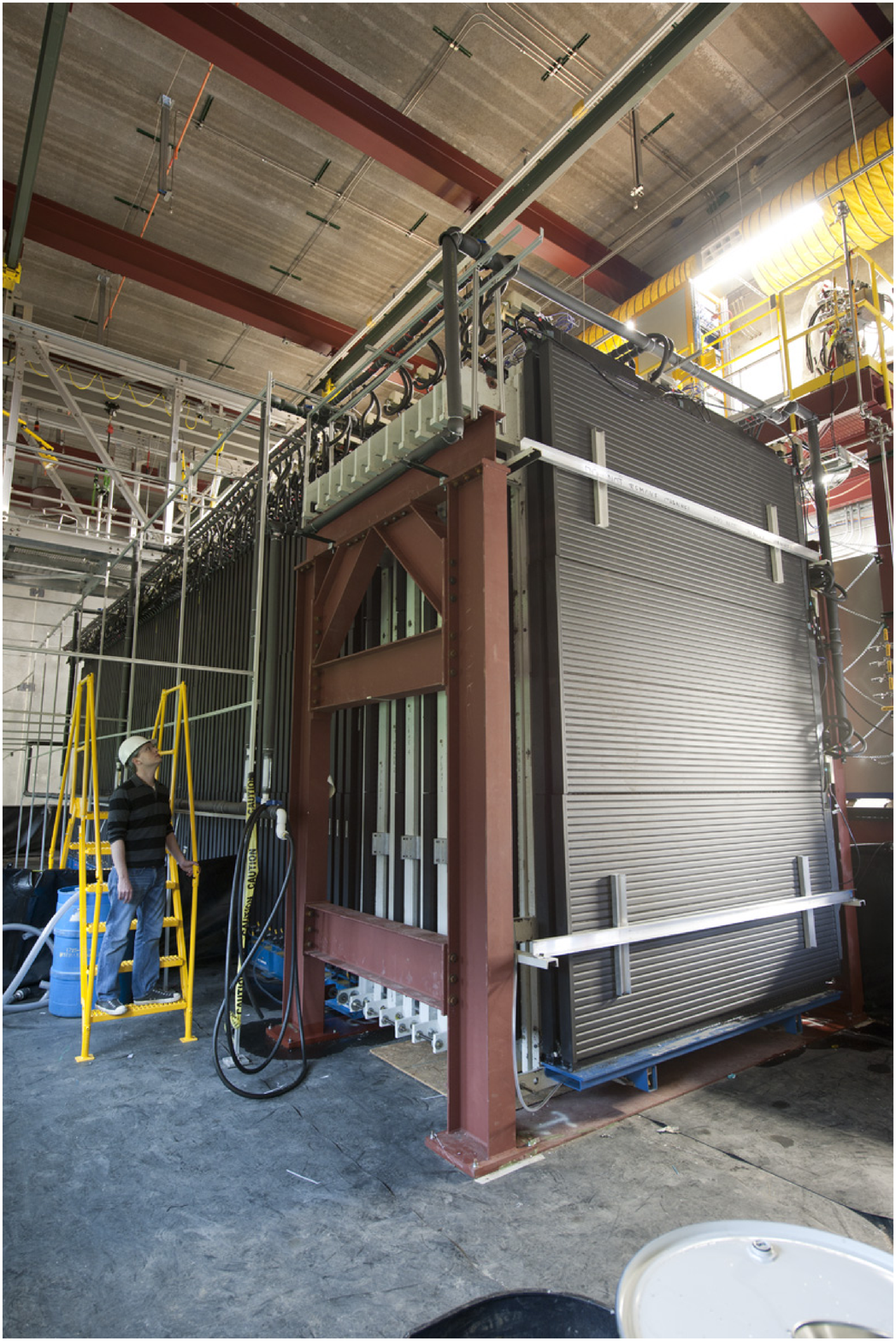}
\caption{Left: Detector components, the near detector and a cell is shown, the dimensions of the detector are 2.9 m wide, 4.2 m high, and 14.3 m long. Center: Detector side view. Right: Prototype detector view.} \label{detectorfigure}
\end{figure*}

The NO$\nu$A Near Detector prototype is located on the surface at Fermilab.
The dimensions of the detector are 2.9 m wide, 4.2 m high, and 14.3 m long. The detector is made of 60$\%$ active PVC plus liquid scintillator and a muon catcher made from iron. The total mass is 222 tons with 125 tons of active material and a fiducial mass for neutrinos interactions of 20 tons. ~Figure~\ref{detectorfigure} illustrates these regions with different colors. The detector is constructed with rigid PVC cells. These cells are filled with liquid scintillator. Each cell has a wavelength shifting fiber; as shown in ~Figure~\ref{detectorfigure} which illustrates a cell with a fiber that is looped at the bottom. When a charged particle goes through the cell, the scintillator produces light. This light bounces around the rectangular cell which has a width of 3.87 cm and depth of 6 cm until the lights gets captured by a wavelength shifting fiber or absorbed by the PVC or the scintillator. Both ends of the fiber are connected to one pixel on an Avalanche Photodiode (APD). This light is converted to an electric signal and processed through the data acquisition system. This prototype provides essential information on; assembly technique, scintillator filling, light yield, APD installation and functioning, electronics installation and functioning, DAQ functioning.
\subsection{The NDOS sees neutrinos from 2 different beams, NuMI and Booster}
\begin{figure*}[h]
\centering
\includegraphics[width=90mm]{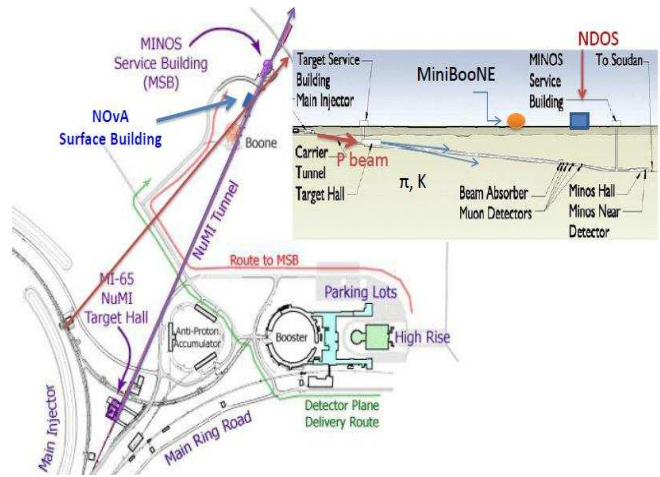}
\caption{The figure on the left illustrates the detector location, the Booster and NuMI neutrino beams are illustrated with the two arrows and the NO$\nu$A surface building is located between the MINOS surface building and MiniBooNE detector hall. The figure on the right shows part of the NuMI underground location, where protons from the NuMI interact with the graphite target and from this interaction mesons are produced.} \label{locationfigure}
\end{figure*}

The NDOS is placed $6.1^{\circ}$ off the axis of the NuMI beam and on the axis of the Booster beam. ~Figure~\ref{locationfigure} illustrates the location of the detector. The detector is located on the surface between the MiniBooNE detector hall and the MINOS service building. The beam of neutrinos from NuMI is generated by focusing 120 GeV protons from the Main Injector onto a graphite target~\cite{NuMI}. This interaction produces mesons (pions and kaons), which are focused by two magnetic focusing horns located downstream the target. The mesons decay in a pipe filled with He at 1atm, which is 675 m in length and 2 m in diameter producing neutrinos. The two magnetic horns are pulsed with a 200kA current. Because the two horns focus positively charged particles and defocus negatively charged particles, changing the horn current polarity produces either a neutrino or an antineutrino beam. The mesons will primarily decay through the channels $\pi^{\pm}\rightarrow\mu^{\pm}\nu_{\mu}(\bar{\nu}_{\mu})$ and $K^{\pm}\rightarrow\mu^{\pm}\nu_{\mu}(\bar{\nu}_{\mu})$. Also the muons decay and produce ${\mu^{\pm}}\rightarrow\nu_{\mu}(\bar{\nu}_{\mu})+e^{\pm}+\nu_{e}(\bar{\nu}_e)$.\\
The beam of neutrinos from the Booster is generated by focusing 8 GeV protons from the Booster onto a beryllium target. The process to produce neutrinos is similar as for the NuMI beam described previously, but with one horn~\cite{Booster}.
\section{Physics goals for NDOS}

\begin{figure}[ht]
\centering
\includegraphics[width=50mm]{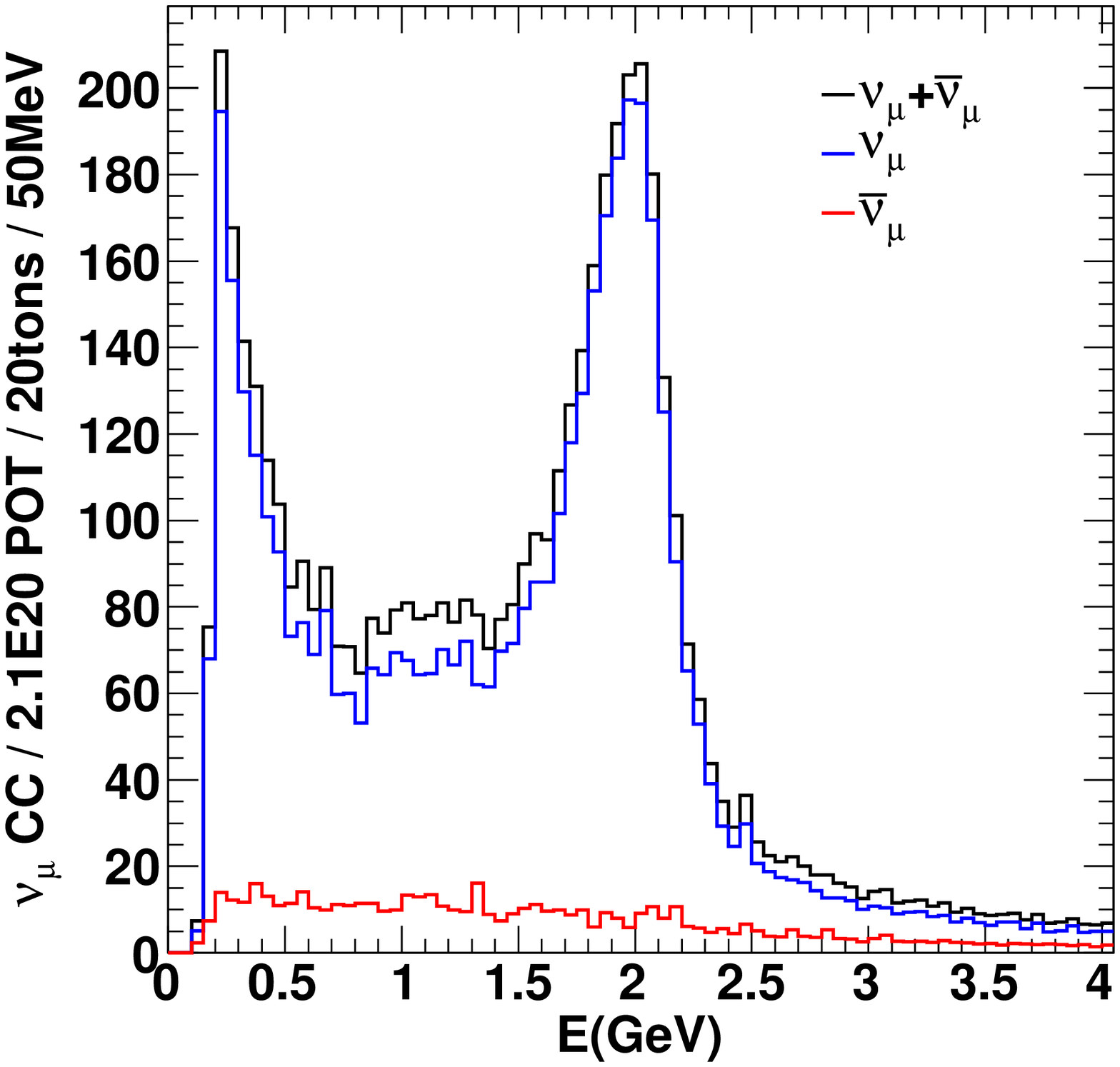}
\includegraphics[width=50mm]{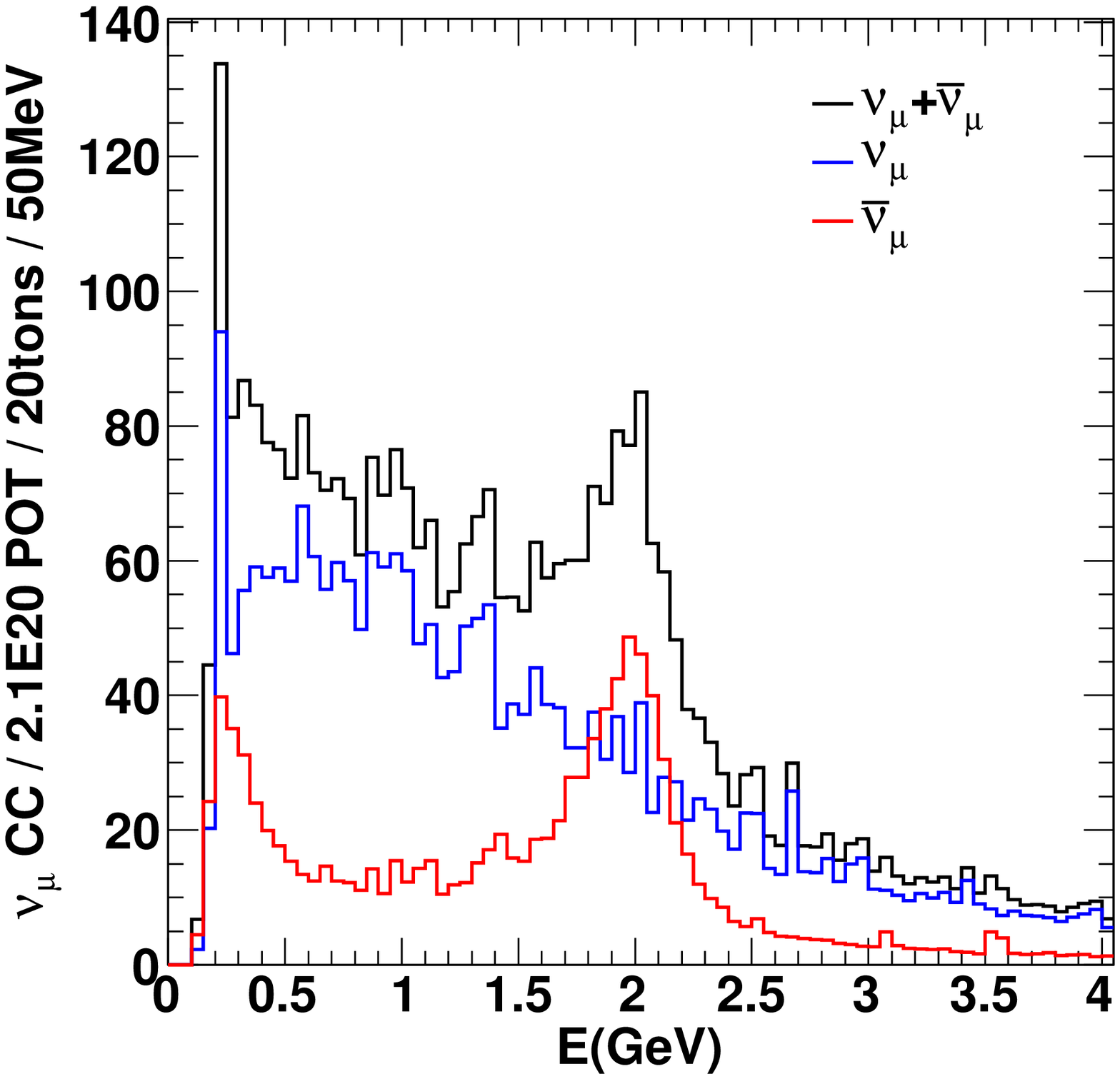} 
\includegraphics[width=50mm]{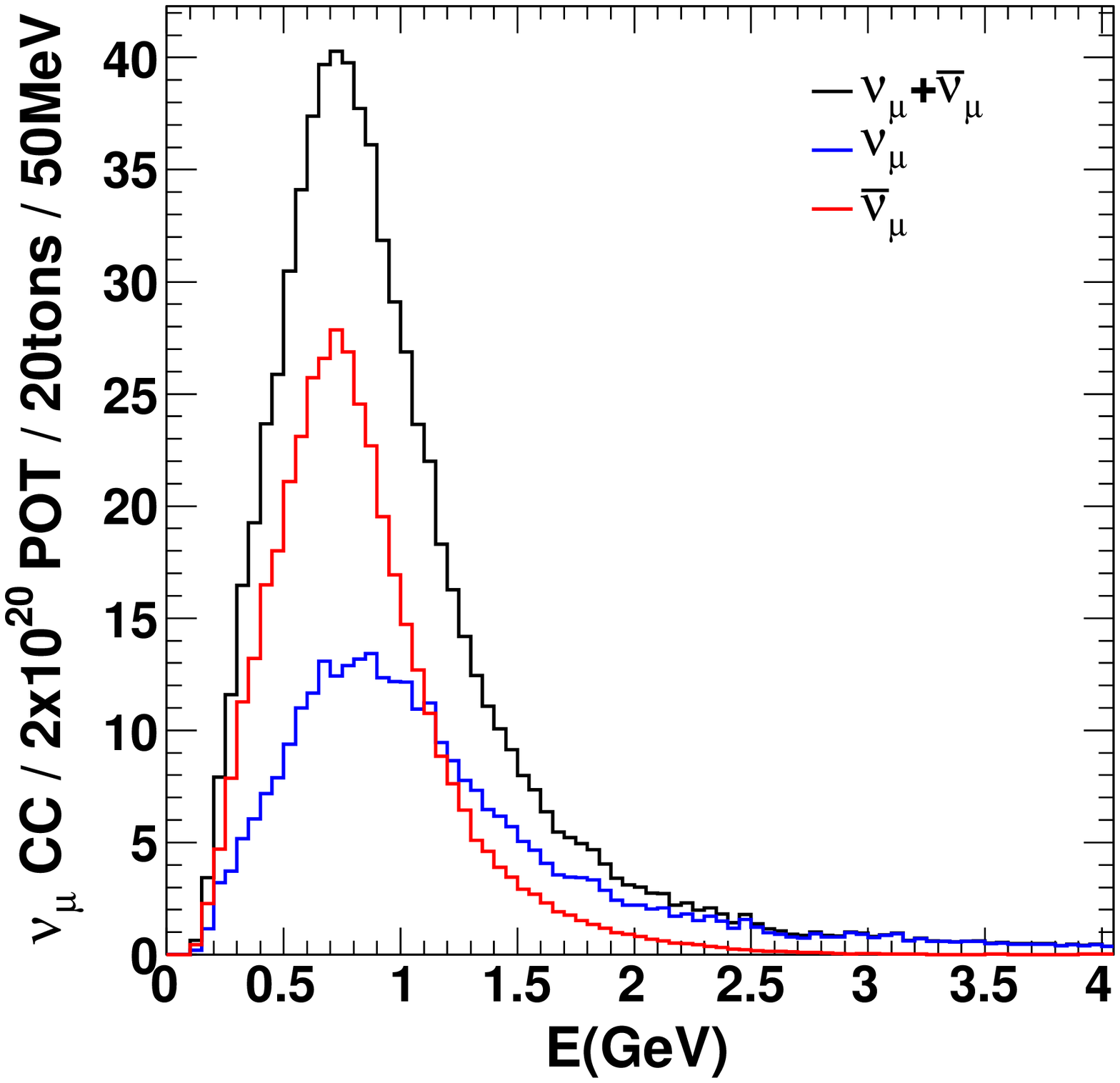}
\caption{Simulated neutrino energy spectra. The energy distribution for neutrinos are shown in blue and for antineutrinos in red. Left: NuMI neutrino beam. Center: NuMI antineutrino beam. Right: Booster antineutrino beam.} \label{spetrumfigure}
\end{figure}
 The NDOS collects neutrino and antineutrino data from NuMI and antineutrino data from the Booster. The simulated neutrino energy spectra are shown in ~Figure~\ref{spetrumfigure}. The neutrino mode from NuMI shows two peaks, the low energy peak is produced from the pion decay and the peak around $2$GeV is produced from the kaon decay.\\
 The physics goals for NDOS are:\\ 
\begin{itemize}
\item Determine the composition of the beam.
\item Investigate the detector sensitivity to cosmic ray background.
\item Study the response of the detector to electron neutrino interactions.
\item Measure the rate of neutrino interactions for the quasi-elastic (QE) interactions.
\end{itemize}

\section{Quasi-elastic Studies in NDOS}
\begin{figure*}[h]
\centering
\includegraphics[width=80mm, height=40mm]{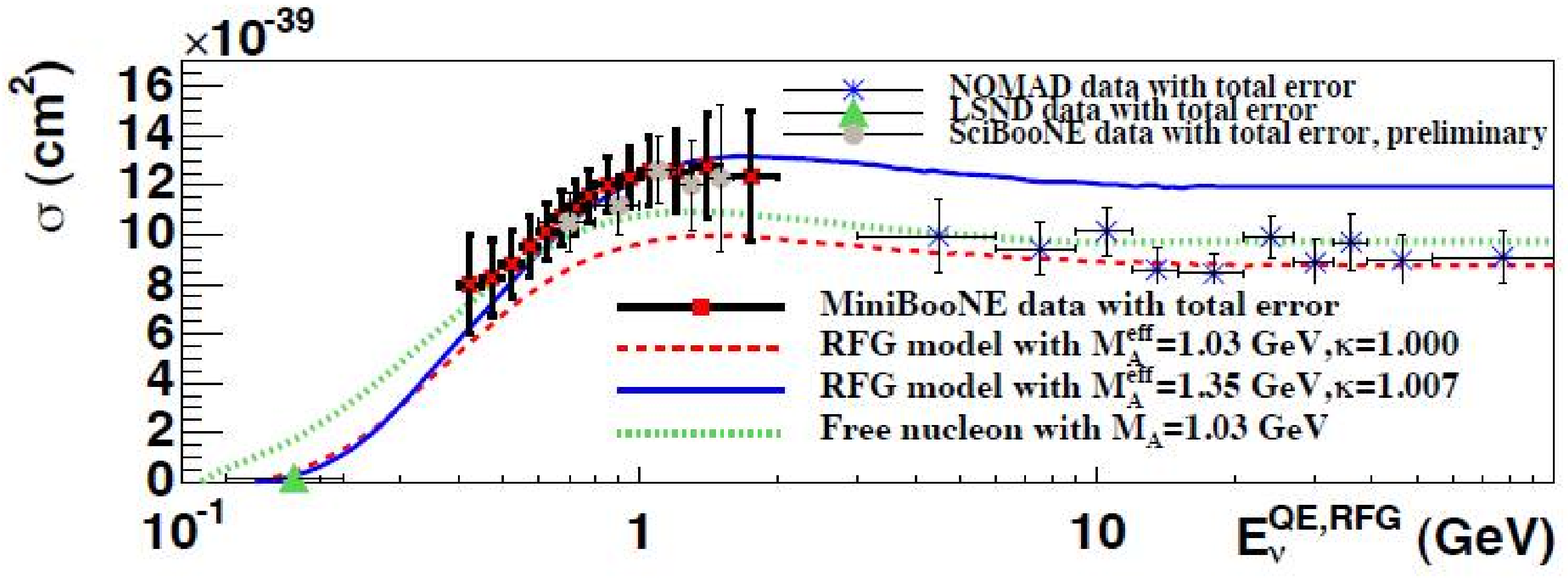}
\caption{Charged current QE cross-section versus neutrino energy.~\cite{thrid}}\label{QEfigure}
\end{figure*}

Recent experiments have made measurements of cross sections for the QE interactions. The MiniBooNE experiment produces neutrinos with a mean energy of approximately $788$MeV ~\cite{first} and the NOMAD experiment uses an average energy of the incoming neutrino around $25.9$GeV~\cite{second}. In addition the SciBooNE experiment using the same neutrino beam as MiniBooNE made a measurement for the low energy region similar to the low energy region of the MiniBooNE experiment. The measurement from SciBooNE experiment is in agreement with the MiniBooNE experiment measurement. ~Figure~\ref{QEfigure} shows the measurements of the charged-current QE cross section from the current experiments. These experiments have covered energy regions above and below $2$GeV. However, around the 2GeV peak these experiments show a disagreement.  The mean $\nu_{\mu}$ energy for the NDOS peaks around 2GeV, which allows us to study this region. The QE studies in NDOS will also help to develop the analysis for the $\theta_{23}$ and $\Delta{m}^2_{32}$ NO$\nu$A measurements.\\
The Near Detector prototype will collect data for a year. For an exposure of $2\times{10}^{20}$ protons on target and 20 tons of fiducial mass, the expected events rates are shown in ~Table~\ref{eventrate}. 
\begin{table}[h]
\begin{center}
\caption{Event Rates in NDOS.}
\begin{tabular}{|l|c|c|c|}
\hline \textbf{21020POT, 20 tons} & \textbf{NuMI Neutrino} & \textbf{NuMI Anti-Neutrino} &
\textbf{Booster Anti-Neutrino}
\\
\hline $\nu_{\mu}+\bar{\nu}_{\mu}$ CC& 4500 & 3300 & 735 \\

\hline In 2 GeV peak & 1500 & 800 &  \\
\hline $\nu_{e}+\bar{\nu}_{e}$ CC & 200 & 160 & 10 \\
\hline NC & 2000 & 1600 & 392 \\
\hline
\end{tabular}
\label{eventrate}
\end{center}
\end{table}

\section{Early Data from NDOS}
\begin{figure*}[h]
\centering
\includegraphics[width=80mm]{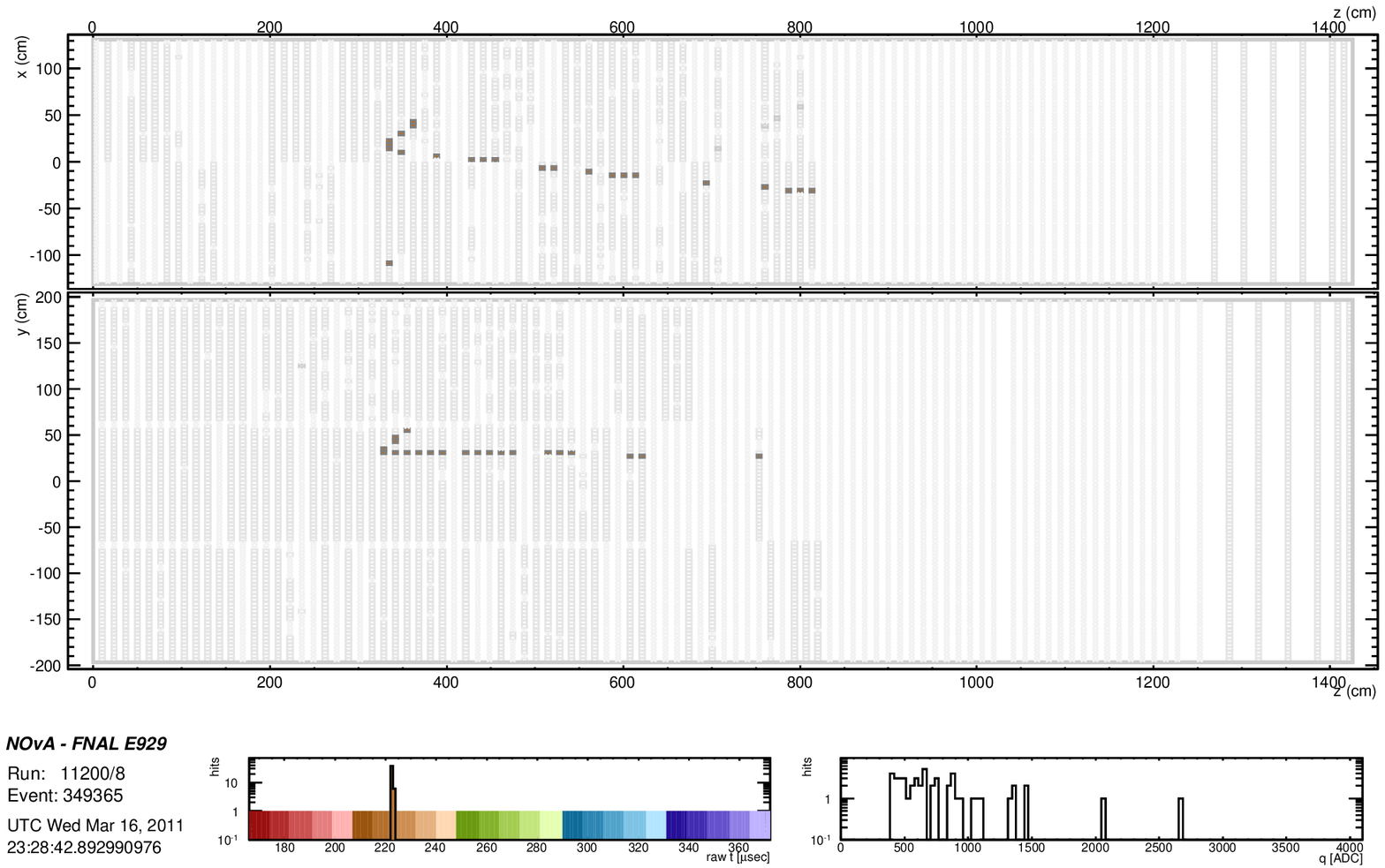}
\includegraphics[width=80mm]{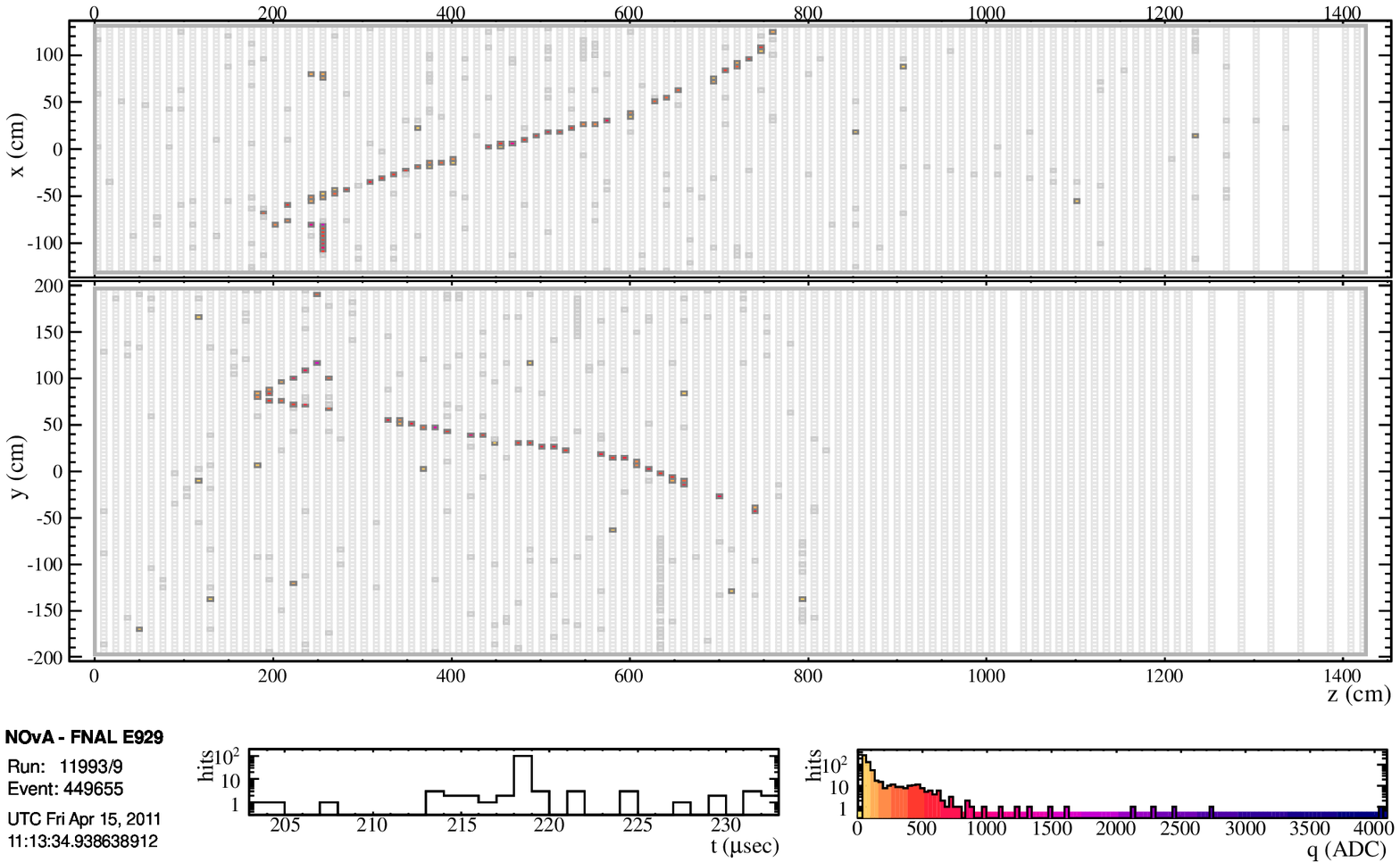}
\caption{Two event displays of neutrino events collected from NuMI beam. The top and bottom in each event display shows the horizontal and the vertical detector view. Left: $\nu_{\mu}$ CC QE candidate. Right: $\nu_{\mu}$ CC candidate} \label{eventsfigure}
\end{figure*}
The NDOS has been collecting neutrino data since December 2010. ~Figure~\ref{eventsfigure} shows two event displays from early data during commissioning at NDOS. The event displays show the two detector views, the top is the horizontal detector view and the bottom is the vertical detector view. The left-hand-side plot shows a $\nu_{\mu}$ charged-current quasi-elastic candidate. This event has two tracks, a long track associated with a muon and a short track associated with a proton. The right-hand side figure shows a charged-current candidate, also an event with two tracks, the long track associated with a muon and the short probably a pion, since the short track has different energy deposition along the track and and some activity at the end of the track. The event displays also show white regions and other regions where the tracks are not visible. Those regions are not instrumented with electronics or are masked because they are noisy.
\subsection{Neutrino Signal in NDOS}
\begin{figure*}[ht]
\centering
\includegraphics[width=50mm,height=45mm]{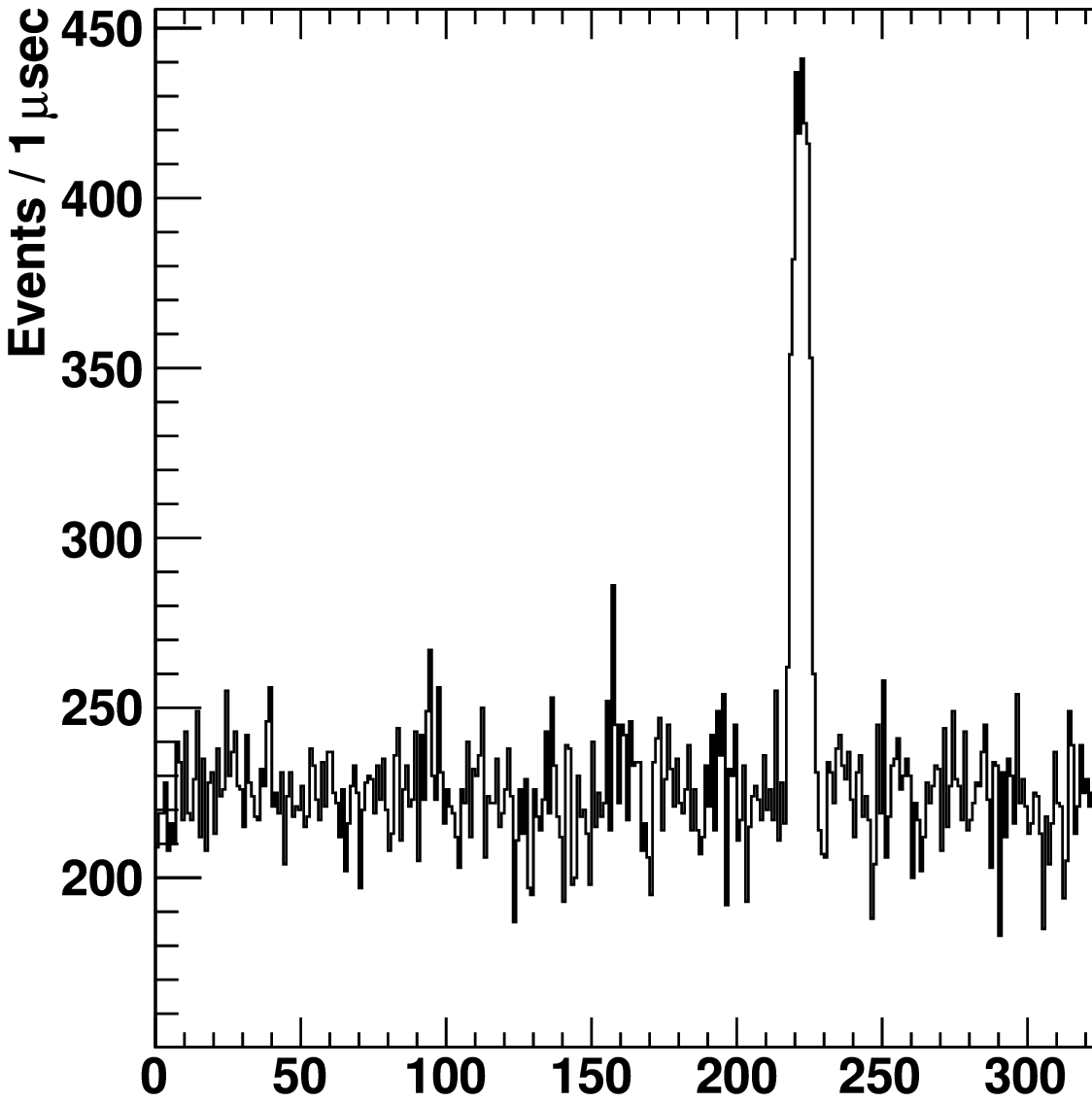}
\includegraphics[width=50mm,height=45mm]{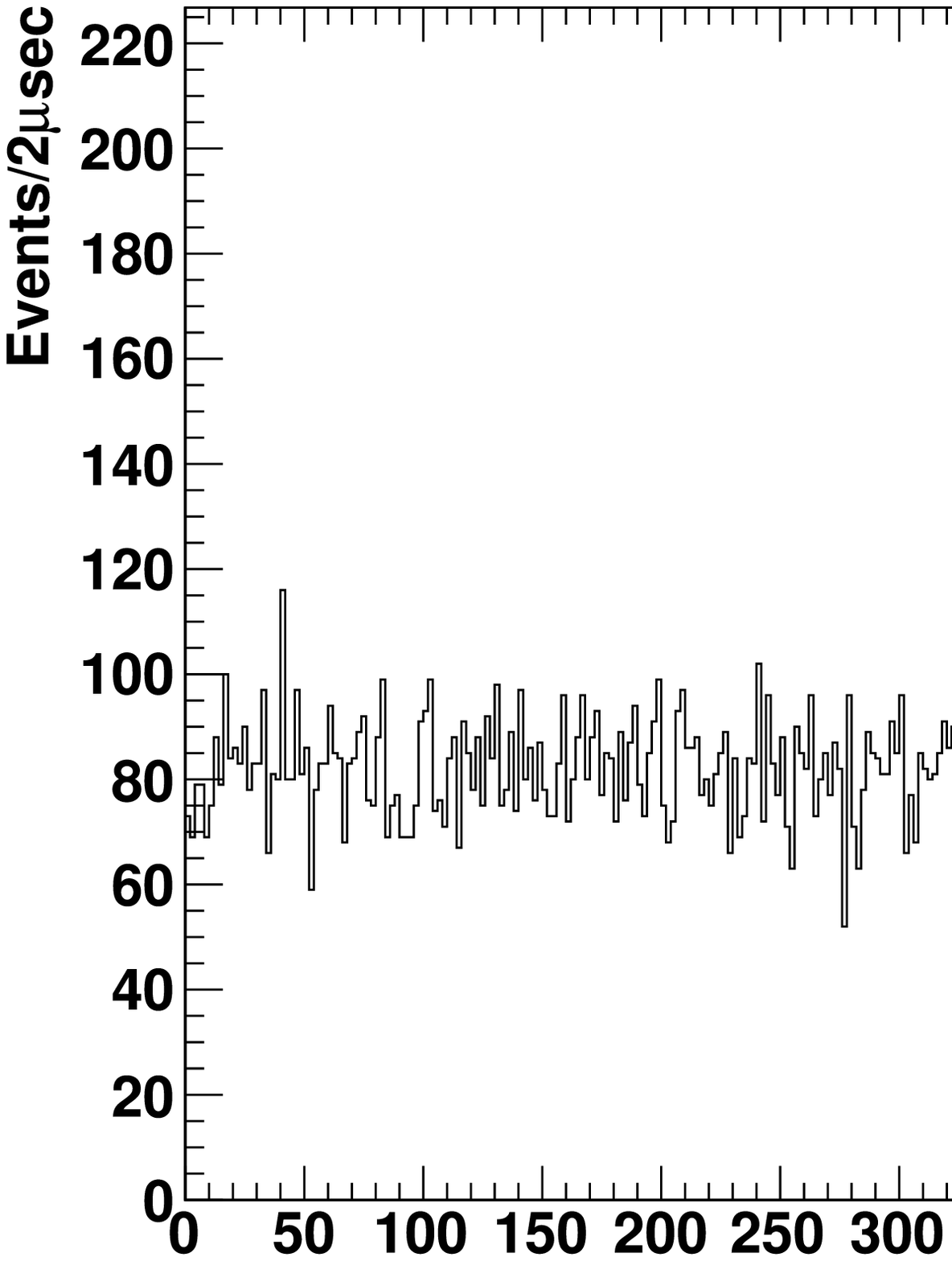}
\caption{Left: Time distribution for NuMI Neutrinos. Right: Time distribution for Booster Neutrinos.} \label{timefigure}
\end{figure*}

\begin{figure*}[ht]
\centering
\includegraphics[width=50mm]{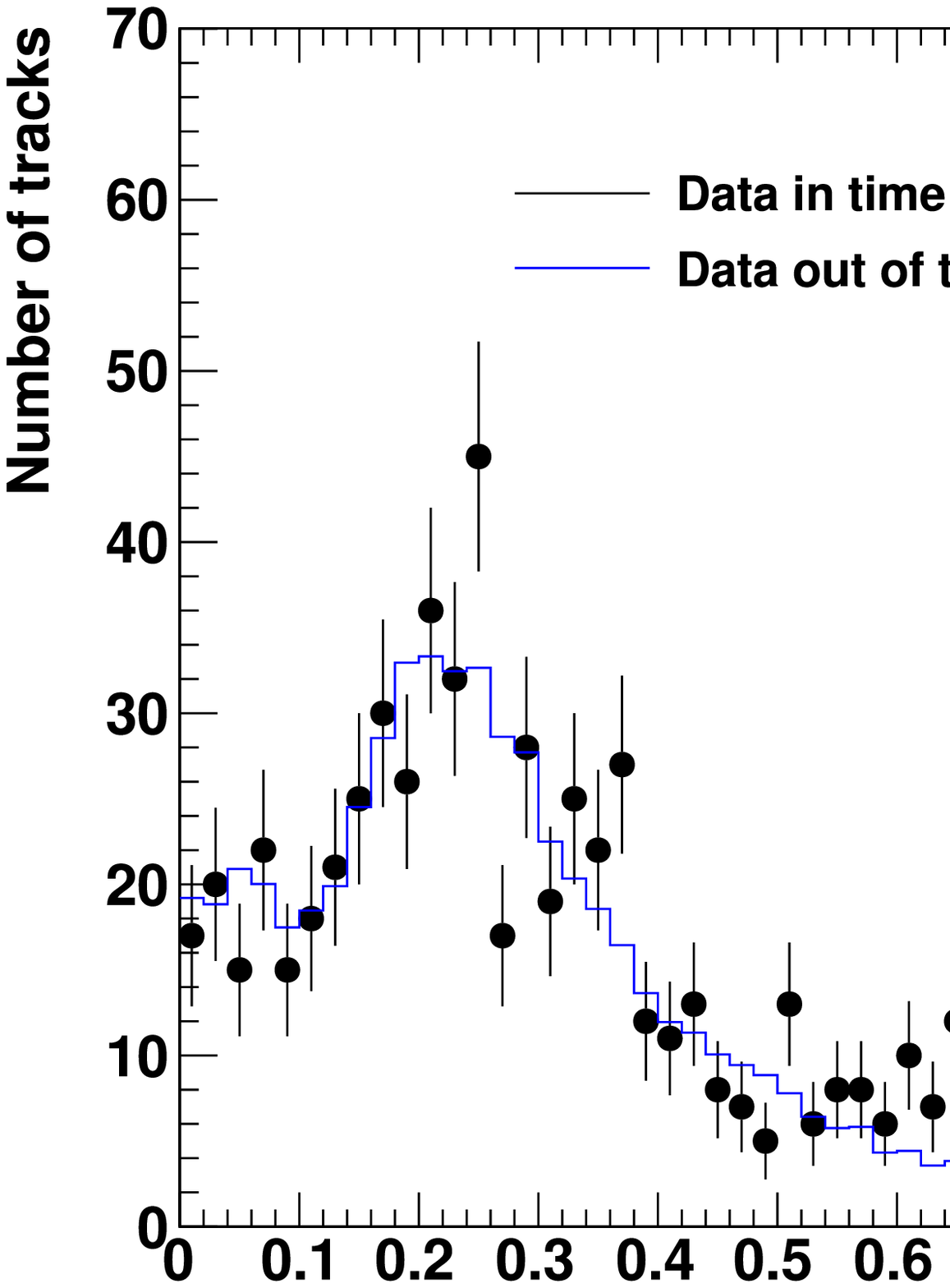}
\includegraphics[width=50mm]{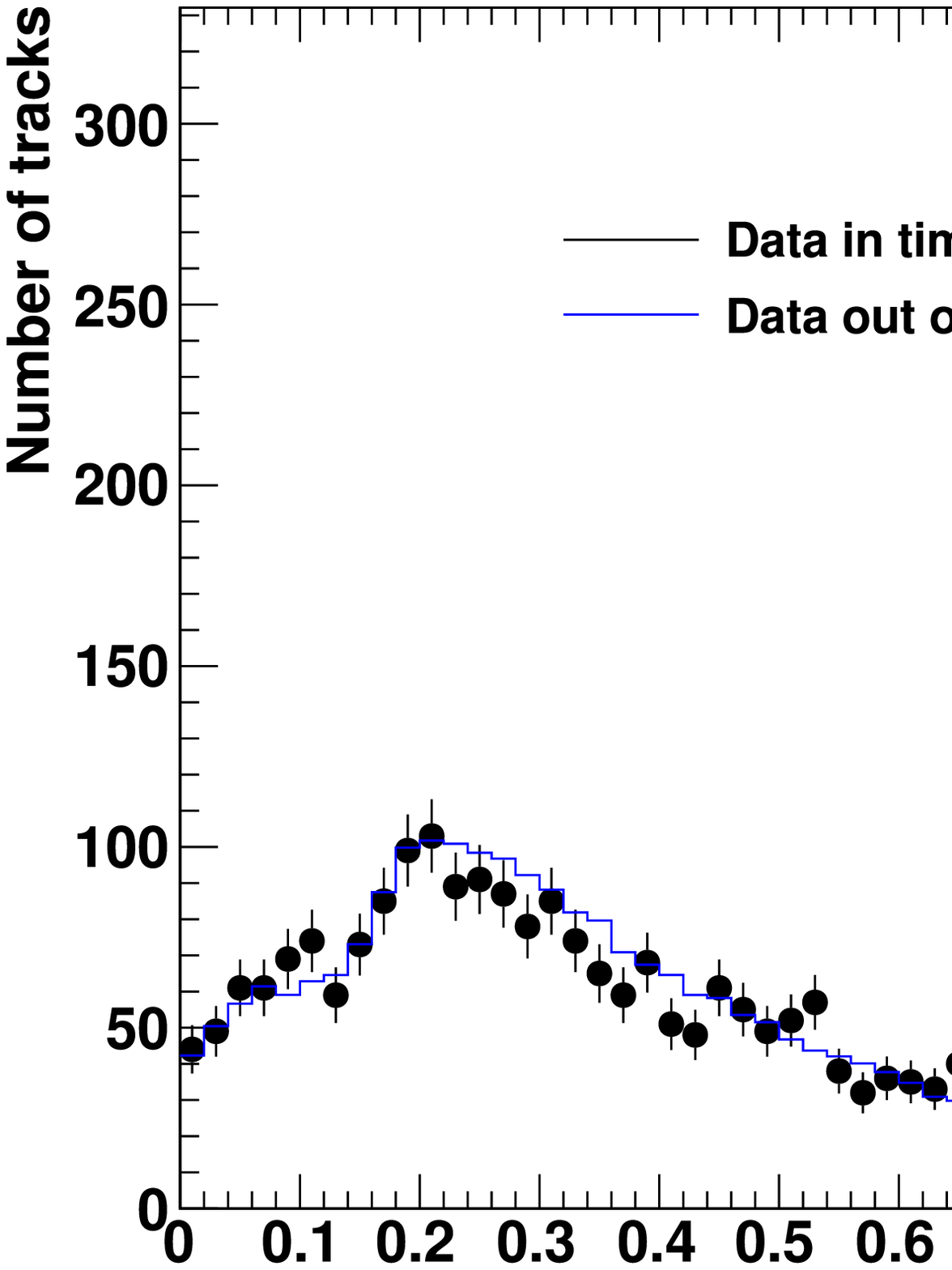}
\caption{Reconstructed particle track angles with respect to the beam direction for the NuMI data. Left: NuMI antineutrino data. Right: NuMI neutrino data.} \label{anglefigure}
\end{figure*}

\begin{figure*}[ht]
\centering
\includegraphics[width=50mm]{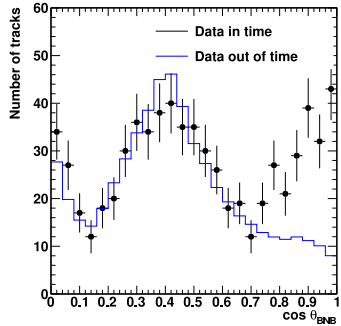}
\caption{Reconstructed particle track angles with respect to the beam direction for the Booster data~\cite{angleBooster}.} \label{anglebnbfigure}
\end{figure*}

 We collect data with a trigger window associated with NuMI and Booster beams. The trigger window is open for 500 $\mu{s}$ synchronized with the 10 $\mu{s}$ long NuMI spill or the 2 $\mu{s}$ Booster spill. ~Figure~\ref{timefigure} shows the time for both beams relative to the start of 500 $\mu{s}$ window. The left plot is the time distribution for the NuMI neutrino interaction candidates and the right is the time distribution for the Booster neutrino interaction candidates. At this stage in the analysis, neutrino interaction candidates required only events which had more than 4 hits in each view and the vertex of the event in the fiducial region. The fiducial region is defined as follows: $|x|<110$ cm, ${y<140}$ cm and the longitudinal coordinate $z>50$ cm and $z<770$ cm.\\ 
Using the fiducial selection and a time cut to select only the events that are in time with the neutrino beam, the reconstructed particle track angle with respect to the beam direction are shown in ~Figure~\ref{anglefigure}. This figure shows the reconstructed angle for both neutrino configurations. The left shows the NuMI neutrino data and the right the NuMI antineutrino data. The distributions show that the neutrino events are peaked around $cos\theta_{NuMI}$ equal 1 and the peak for small values of $cos\theta_{NuMI}$ is due to cosmic background. The solid line distribution shows the data not in the NuMI spill window. 
We have collected $5.6\times{10}^{19}$ protons on target (POT) during the antineutrino run, yielding 1001 antineutrino NuMI events with 69 cosmic background and for the neutrino run we have collected $8.4\times{10}^{18}$ POT, yielding 253 neutrino NuMI events with 39 cosmic background. ~Figure~\ref{anglebnbfigure} shows the reconstructed particle track angle with respect to the beam direction for the Booster neutrinos. This plot shows that events above $cos\theta_{BNB}>0.7$ are neutrino events and the solid line distribution is the data outside the Booster spill window.
\subsection{Track Length for NuMI signal events}
\begin{figure*}[ht]
\centering
\includegraphics[width=60mm,height=45mm]{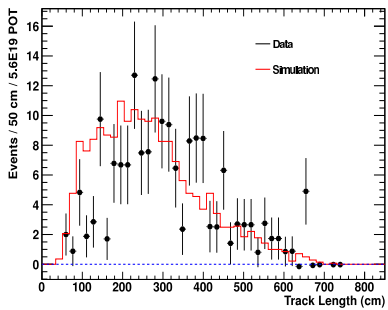}
\caption{Track length of the fully contained events for the antineutrino data.} \label{trackfigure}
\end{figure*}

~Figure~\ref{trackfigure} shows the distribution for the track length of the fully contained neutrino events from the antineutrino run. These fully contained events have the vertex and the end of the track inside the fiducial volume. The cosmic background has been subtracted from the data. The solid line is the MC simulation normalized to data exposure. This MC contains all the interactions charged-current and neutral current interactions. ~Table~\ref{datamctable} shows a comparison of data-MC for the events in the fiducial region and for the fully contained events. The events in the fiducial region show a discrepancy. Scanning the events showed that the discrepancy results from the absence in the MC simulation of muons originating from neutrinos interactions in the rock beneath of the detector. Otherwise the fully contained events display reasonable data-MC agreement. 
\begin{table}[ht]
\begin{center}
\caption{Data-MC comparisons for neutrino events}
\begin{tabular}{|l|c|c|c|}
\hline  & \textbf{NuMI Data} & \textbf{Cosmic Bg} &
\textbf{MC Simulation}\\
\hline Fiducial&1001 & 69 & 861 \\
\hline Fully Contained&184 & 12 & 187 \\
\hline
\end{tabular}
\label{datamctable}
\end{center}
\end{table}

In summary we have developed a selection criterion to find neutrino candidates and separate them from cosmic background. 
\section{Charged Current studies in NDOS}
\begin{figure*}[ht]
\centering
\includegraphics[width=80mm]{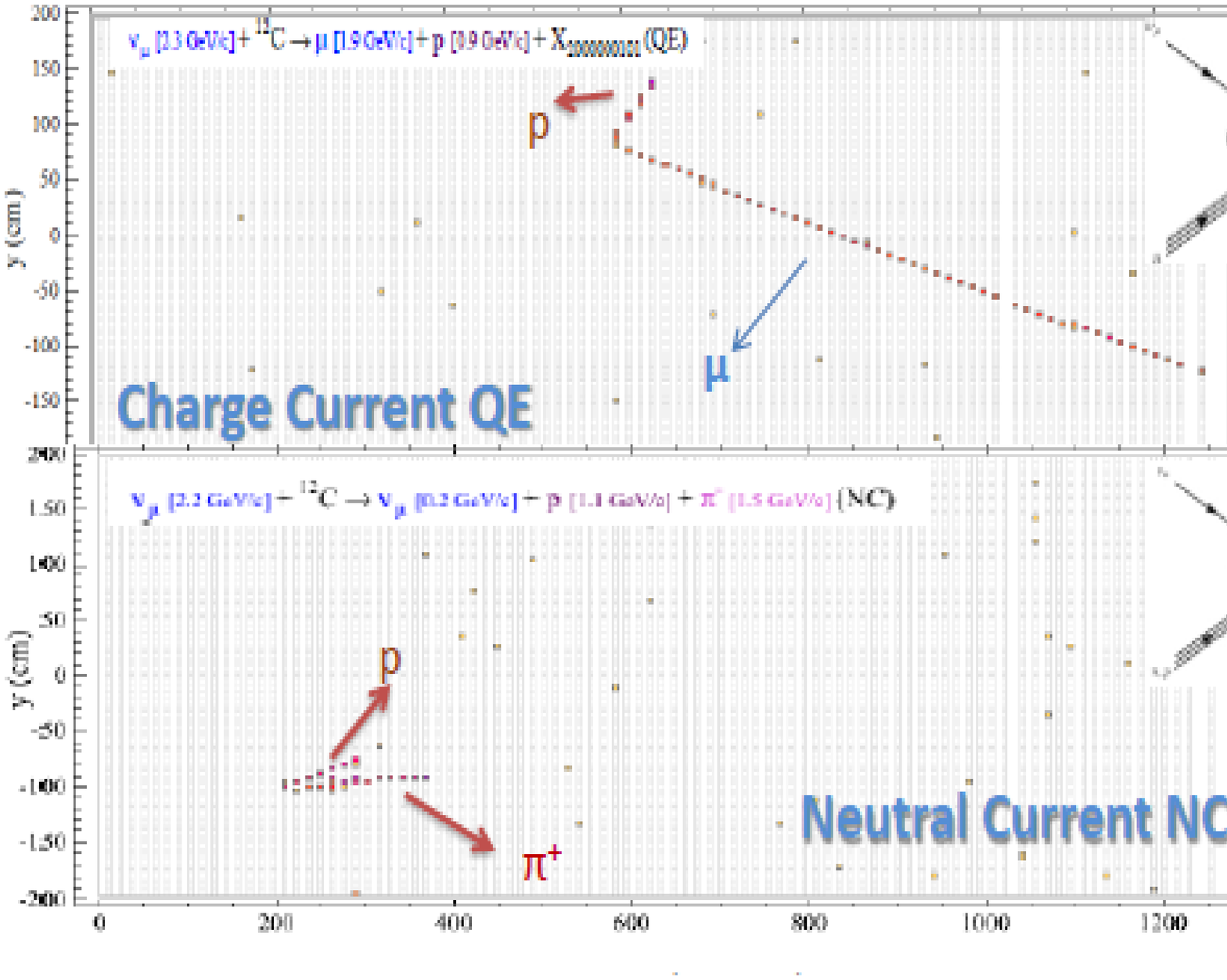}
\caption{Top: Charged-current quasi-elastic interaction. Bottom: Neutral-current interaction} \label{mceventfigure}
\end{figure*}

The next step is to positively identify muon neutrino charged-current events. ~Figure~\ref{mceventfigure} on the top plot shows an example of a simulated charged-current quasi-elastic interactions and the bottom plot shows a simulated neutral current interaction. Both interactions contain two tracks. The QE interaction has a track from a muon and a track from a proton. The muon has a long trajectory and the proton a short trajectory. The proton deposits more energy per cell than the muon. The neutral current interaction has a proton and pion, the pion from the neutral current interaction deposits more energy per plane than the muon for the quasi elastic interaction. The energy deposition and distance traveled by the pion and the muon are different. This allows us to separate charged-current interactions from neutral current interactions. 

\subsection{Selection Method}
\begin{figure*}[ht]
\centering
\includegraphics[width=50mm]{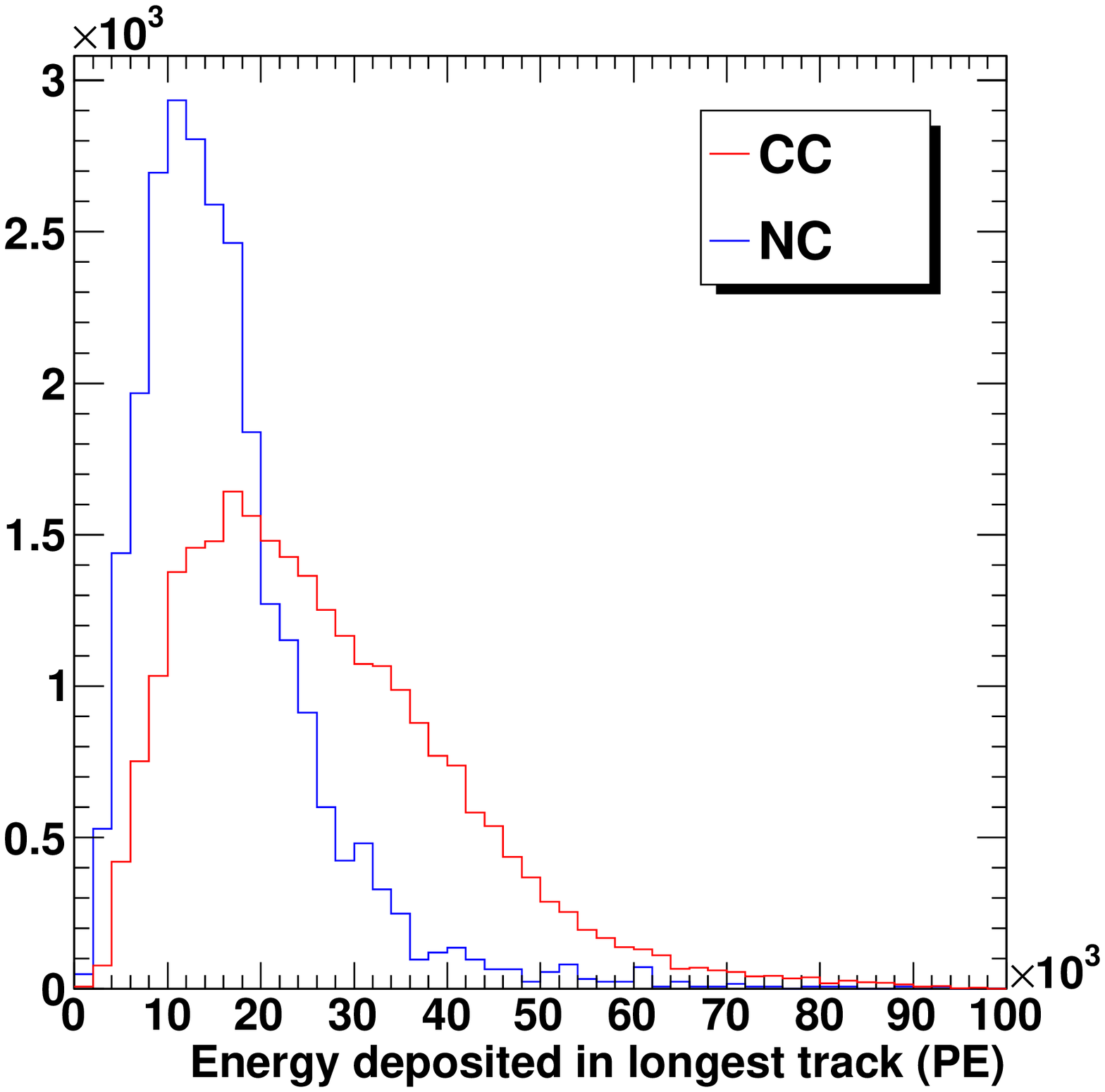}
\includegraphics[width=50mm]{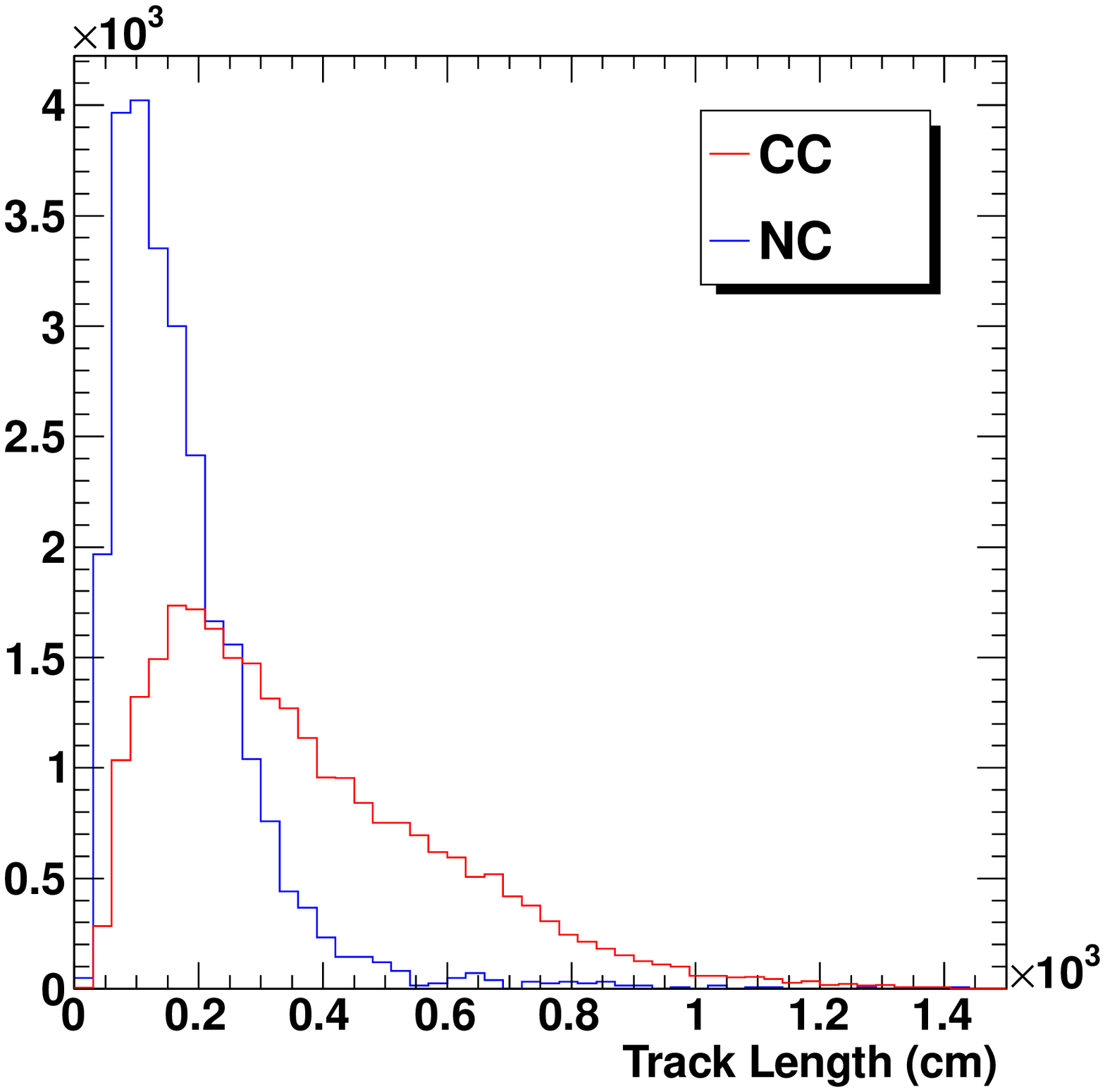}
\includegraphics[width=50mm]{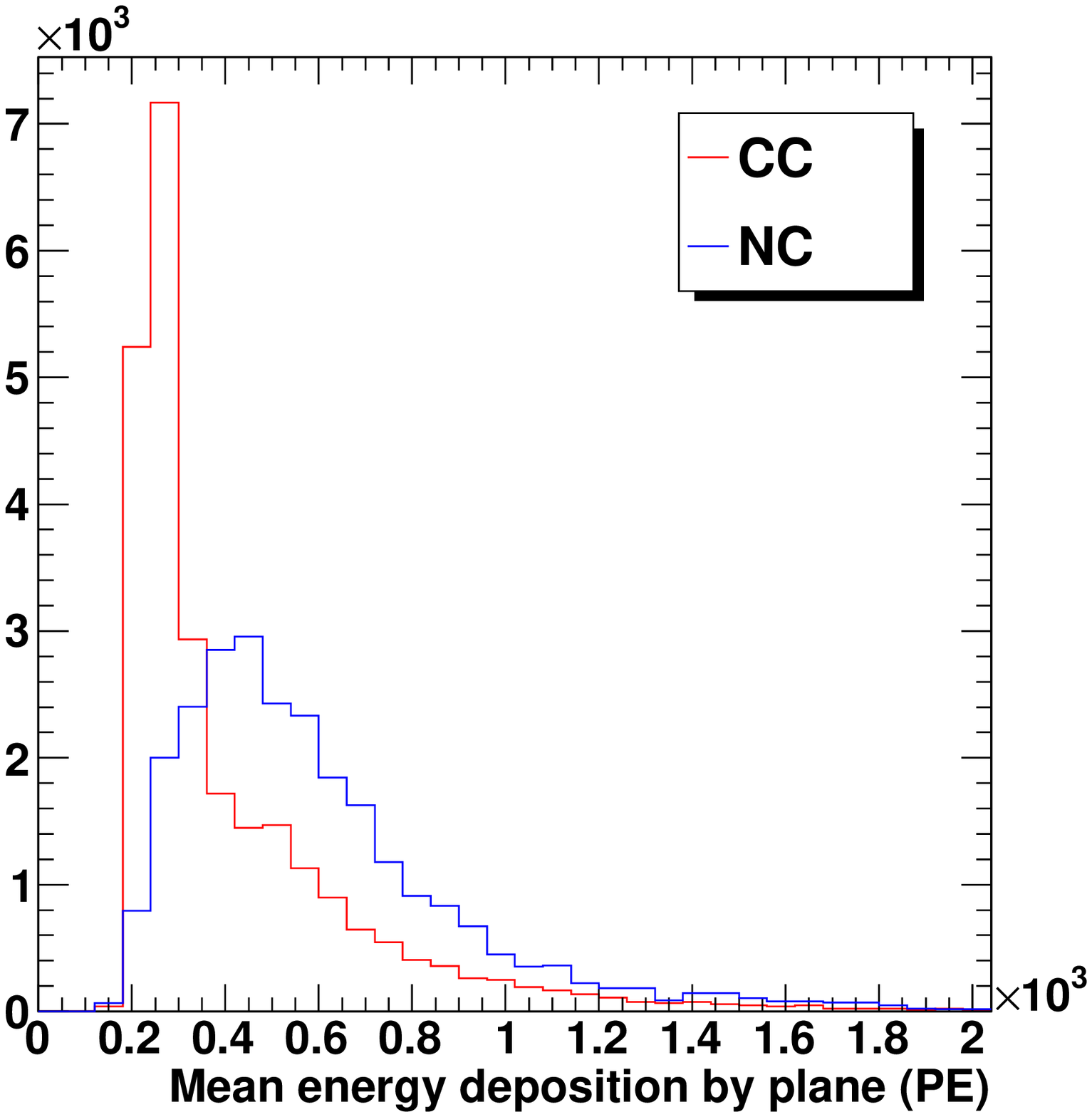}
\caption{Left: Energy deposited in longest track (PE). Center: Track Length (cm). Right: Mean energy deposition by plane (PE).  }\label{variablesfigure}
\end{figure*}

We are currently developing a selection method to select charged-current CC interactions from neutral-current NC interactions. The method is based on a simple likelihood method which uses as input three probability density functions (PDFs).\\
We build a probability given by the multiplication of each PDF for each CC and NC interaction. The probability can be written as follows:
\begin{equation}
P_{CC,NC}={\mathop{\Pi}_{i=1}^3}f_i(x_i)_{CC,NC},
\end{equation}
where $f_i(x_i)$ are the individual PDFs for the CC and the NC interactions.\\
Using these probabilities an event separation is defined:
\begin{equation}
S=(ln(P_{CC})-ln(P_{NC})).
\label{separation}
\end{equation}
Using the antineutrino mode MC, the distributions of the input variables used for each PDF are obtained and shown in ~Figure~\ref{variablesfigure}. The fiducial selection defined is applied when calculating the three PDFs. In addition to this selection we require 4 hits in each view and the angle direction selection cos$\theta_{NuMI}>0.7$, to reject the cosmic background. The first plot on the left shows the energy deposited by the longest track in photoelectron units (PE). The center plot displays the track length of the longest track, and demonstrates that tracks from CC-induced events are longer than the ones reconstructed in NC interactions. The third plot is the mean energy deposition by plane in unity of PE. This plot shows that NC interactions deposit more energy per plane than CC interactions.
\subsection{Separation}
\begin{figure*}[ht]
\centering
\includegraphics[width=50mm,height=45mm]{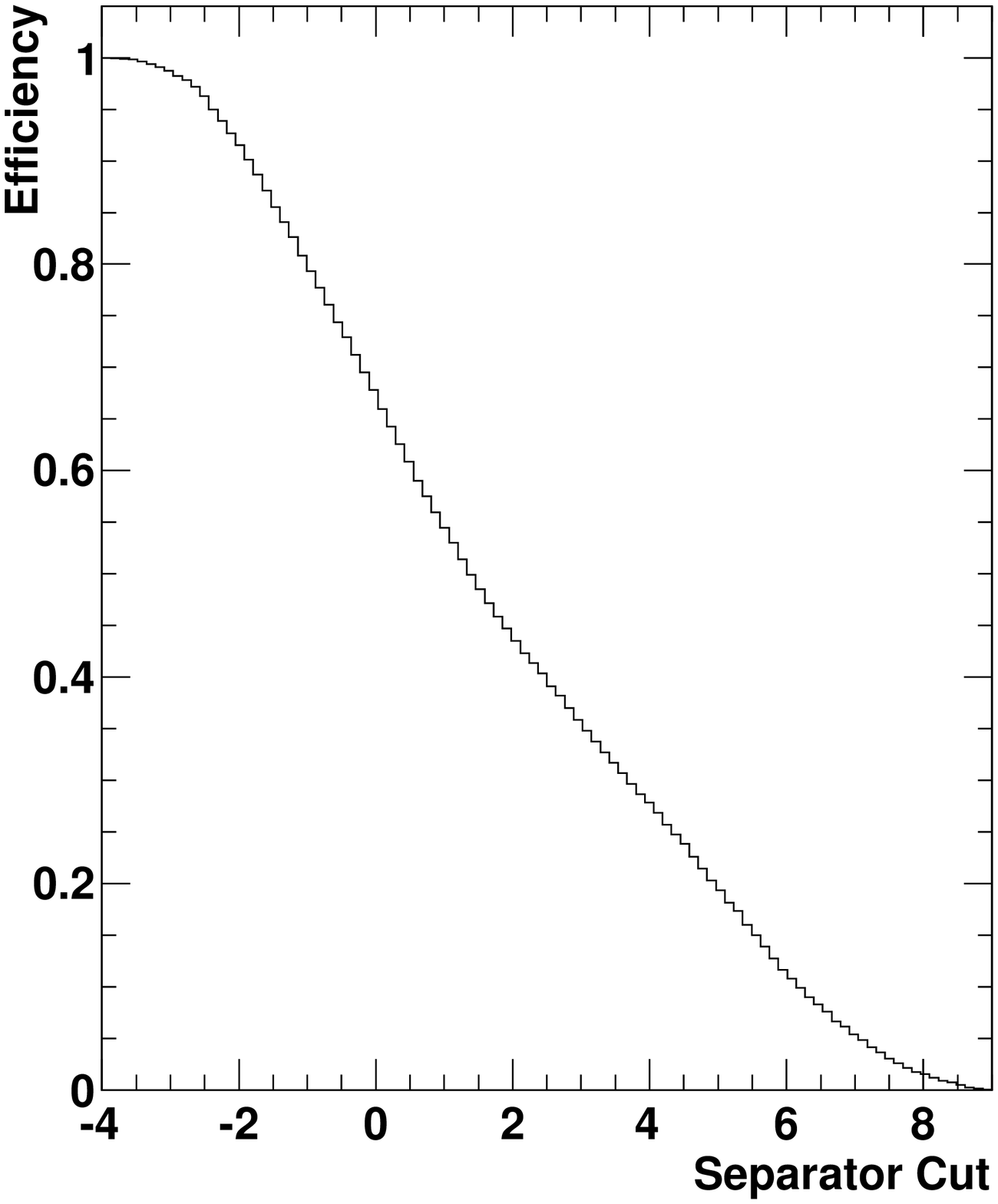}
\includegraphics[width=50mm,height=45mm]{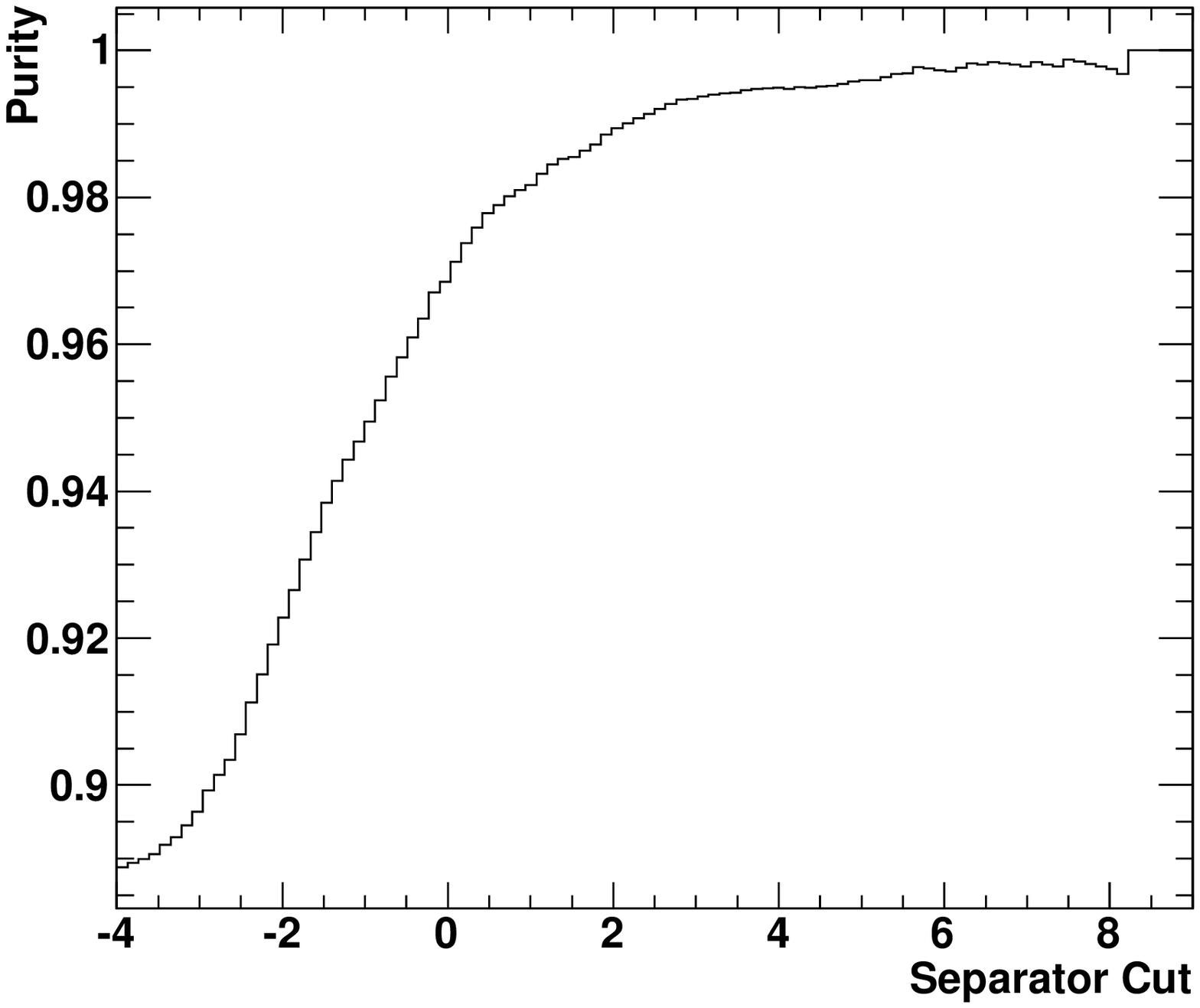}
\caption{Left: Efficiency vs separator cut. Right: Purity vs separator cut.} \label{example_figure16}\label{separationfigure2}
\end{figure*}

Using the formulation in Eq.~\ref{separation}, we compute the simulated CC/NC event separator.\\
\begin{figure*}[ht]
\centering
\includegraphics[width=50mm,height=45mm]{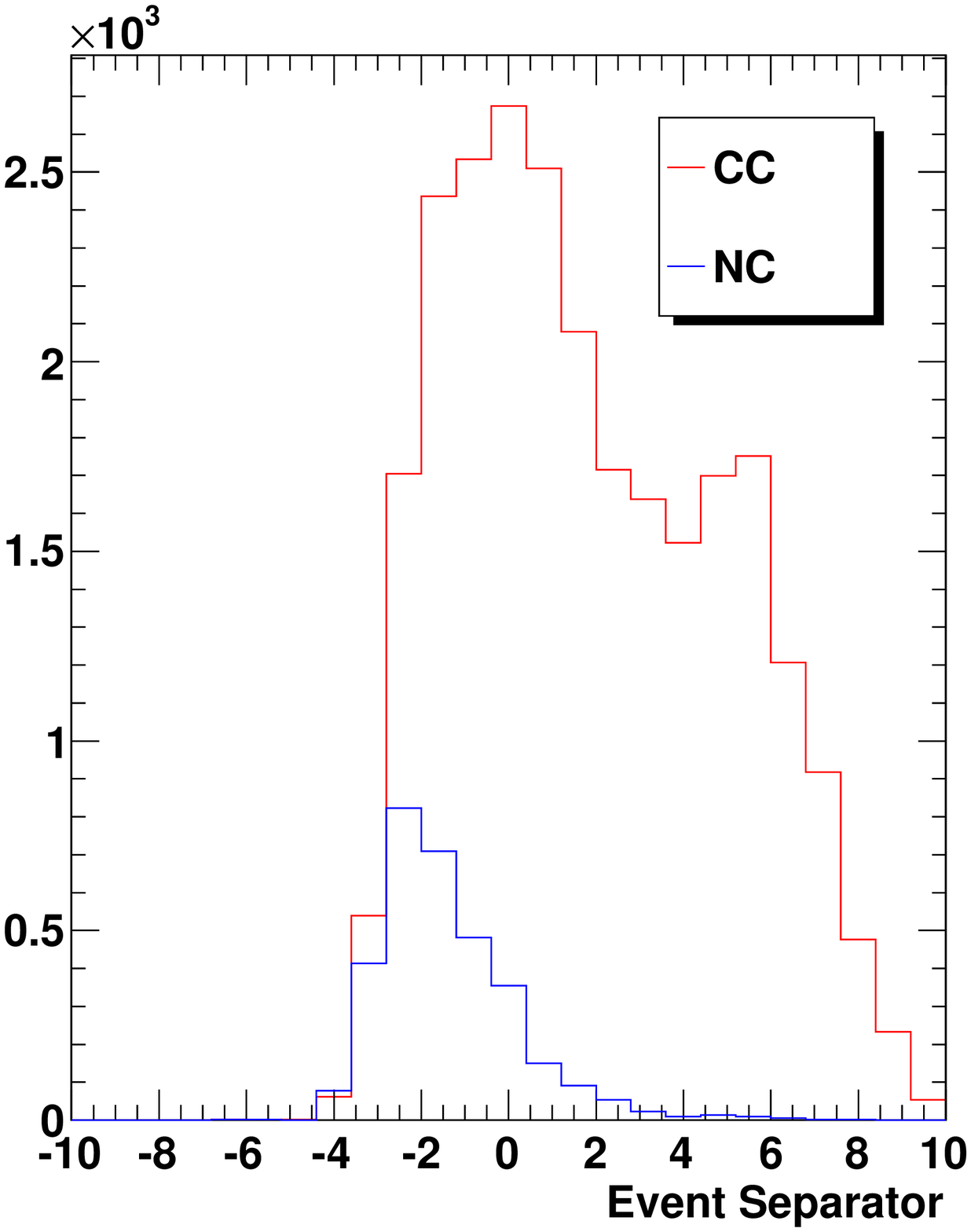}
\includegraphics[width=50mm,height=45mm]{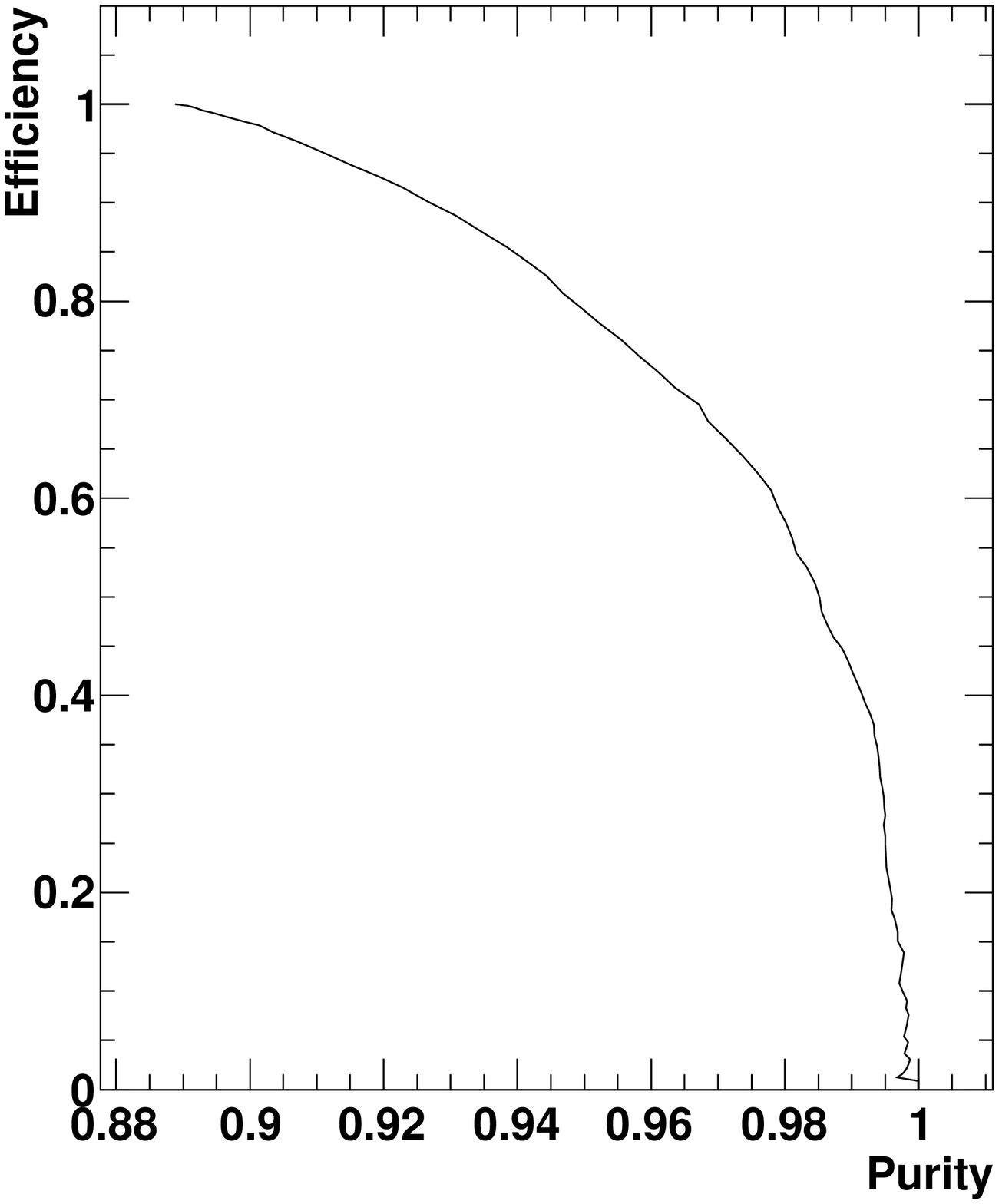}
\caption{Identify simulations. Left: Event Separator. Right: efficiency vs purity} \label{separationfigure}
\end{figure*}
The left distribution in ~Figure~\ref{separationfigure} shows the event separator for the CC and NC interactions and the right distribution the efficiency vs purity. The event separator for the CC interaction shows two peaks. The two peaks are due of the neutrino energy spectrum. This is shown in the simulated neutrino energy spectrum ~Figure~\ref{spetrumfigure}.\\
 The NDOS will collect a small data sample of CC interactions and NC interactions. Therefore, it is essential to tune the selection so that most CC events are retained, but with sufficient background rejection necessary for further analysis.  ~Figure~\ref{separationfigure2} shows the efficiency and purity as a function of each separator cut. A good compromise is obtained by accepting all events with a value of the separator variable larger than 1.5. This selection results in $94\%$ purity and $80\%$ efficiency. \\
An alternative visual scanning method was used to select CC events. This method gives a purity of $92\%$ and an efficiency of $85\%$. The results from the scanning method are comparable with the simple likehood method.\\
\section{Future Work and Conclusions}
The preliminary CC selection method presented here will improve as the quality of the event reconstruction method improves. A possible next step will be to improve the $\pi^{+}$, $\pi^{-}$ rejection using information from Michel electrons, which are produced in end-point decay of muons. This will allow us to differentiate the muons from CC interactions from the pions in NC interactions. At a later stage, identification of quasi-elastic $\nu_{\mu}$ CC interactions will be pursued.\\
NO$\nu$A has a working prototype Near Detector, taking neutrino and cosmic data. The NDOS is taking data from both the NuMI and Booster neutrino beams. Data analysis methods are currently under development, with focus on selection of CC events and later on quasi-elastic CC events. When the Near Detector starts operations underground in 2013, it will collect much higher statistics, which will allow us to make neutrino cross section measurements with high precision.




\bigskip 

\begin{thebibliography}{9}   
\bibitem{zero}
D. S. Ayres el al., The NO$\nu$A technical design report, Fermilab-Design-2007-01(2007).
\bibitem{NuMI} K. Anderson, et al., NuMI facility technical design report, Tech. Rep. Fermilab-Design-1998-01(1998).
\bibitem{Booster}A.A. Aguilar-Arevalo et al., Phys. Rev. D79:072002(2009).
\bibitem{first}
A. A. Aguilar-Arevalo, et al., First Measurement of the Muon Neutrino Charged Current Quasielastic Double Differential Cross Section, Phys. Rev. Lett. 100, 032301(2008),0706.0926.
\bibitem{second}
V. Lyubushlkin et al., A study of quasi-elastic muon neutrino and antineutrino scattering in the NOMAD experiment, arXiv:0812.4543, (2009).
\bibitem{thrid}
 M. O. Wascko, Nuclear Physics B Proceeding Supplement 00, Recent Measurements of Neutrino-Nucleus QuasiElastic Scattering (2011),1-5.
\bibitem{angleBooster}
C. Johnson, Internal NO$\nu$A Document, NO$\nu$A-doc-5873(2011).

\end{thebibliography}

\end{document}